# Nitrogen-Containing Species in the Structure of the Synthesized Hydroxyapatite.

**Marat Gafurov[a], Timur Biktagirov[a], Georgy Mamin[a], Yaroslav Filippov[b], Elena Klimashina[b], Valery Putlayev[b], Boris Yavkin[a], and Sergei Orlinskii[a]**

[a] *Kazan Federal University, Institute of Physics, 420008 Kazan, Russian Federation*
[b] *M.V. Lomonosov State University, Department of Material Science, 119992, Moscow, Russian Federation*

Synthesized by the wet chemical precipitation technique hydroxyapatite powders (HAp) with the sizes of the crystallites of (20-50) nm and 1 μm were analyzed by multi-frequency (9 and 95 GHz) electron paramagnetic resonance (EPR) methods. It is shown that during the synthesis process nitrate anions from the reagents (by-products) could incorporate into the HAp structure. The relaxation times and EPR parameters of the axially symmetric $NO_3^{2-}$ paramagnetic centers detected after X-ray irradiation of the product at room temperature are measured with high accuracy. Assignments of the observed EPR spectra as compared with that given in literature are discussed. Analyses of electron-nuclear double resonance (ENDOR) data from $^1$H and $^{31}$P nuclei and *ab-initio* density functional theory (DFT) calculations allow suggesting that the paramagnetic centers and nitrate anions as the precursors of $NO_3^{2-}$ radicals preferably occupy $PO_4^{3-}$ site in the HAp structure.

## 1. Introduction

Synthetic hydroxyapatite $Ca_{10}(PO_4)_6(OH)_2$ (HAp) is widely used for the variety of applications. Due to its excellent biocompatibility, osteoconductivity and chemical composition mirroring that of bone mineral and teeth enamel it is extensively employed for the hard tissues implantation [1]. It is thought that HAp nanoparticles (nano-HAP) are better candidates for an apatite substitute of bone in biomedical applications than micro-sized hydroxyapatite (micro-HAp) [2]. Nano-HAp has the potential to improve current disease diagnosis due to their ability to circulate in the blood, to deliver and to distribute a payload to image tissues and cells in the body for bio-imaging and therapeutical applications [3, 4]. Antutumor (inhibitory) action of nano-HAp is reported [5]. High sorption activity to a number of various anions and cations including those of some heavy metals and radio nuclides [6] makes the HAp based substances important not only in the biomedical area but also for waste management and in catalysts production [7, 8].

HAp presents a good model of biological apatite only if it is modified by the presence of many other ions. Among the diversity of the methods of the synthesis of the modified HAp powders and nanoparticles the wet (precipitation) chemical procedure involving the aqueous solutions of calcium nitrate, diammonium hydrogen phosphate and ammonium hydroxide is the most commonly in use [9]. The drawback of this fast production rate and low processing costs procedure is that one has to spent time on washing the precipitate to remove nitrate residues [10]. Influence of the nitrogen species on cell metabolism and blood circulatory system and their significance in monitoring of development of various diseases is widely recognized though is still hotly debated [11]. Low concentrations of the nitrogen-containing components appear to be necessary for bone resorption while their high levels have an inhibitory effect [12].

Size effects and the biological effects of the nanoparticles are not yet fully understood [13]. Transition to the nanometer scale could lead to the distinctly different properties of materials. Among other things the doping process becomes a challenging task, the positions and structures of the dopants even incorporated into the nanostructure could vary with the size significantly and depend on the synthesis method and conditions to a far greater degree than that in a bulk material. The introduced dopants, in their turn, can distort the local surrounding that cannot be totally compensated in the microscopic scale due to the restricted size. All these make complicated the task to synthesize nano-materials with the desired parameters and forces to apply some additional (unusual) analytical tools for their characterization.

To investigate the structure and local environments of the impurities wide ranges of techniques are applied [6, 14]: X-ray diffraction (XRD), nuclear magnetic resonance (NMR), infrared (IR) and Raman spectroscopy, *etc*. Methods of electron paramagnetic/spin resonance (EPR/ESR) are well known as the most sensitive tools for the detection, identification and quantification of the naturally existing or artificially created radicals in liquids and solids [15]. Conventional continuous-wave (cw) X-band EPR (frequency of about 9 GHz) and ENDOR (electron-nuclear double resonance) experiments which combine the abilities of NMR with a high sensitivity of EPR are widely used for investigations of the radiation defects in apatite like materials. The lasts are usually ascribed to the different carbonate, phosphorous, hydroxyl and oxygen radicals (see [16] for the recent review). Increased sensitivity and selectivity of modern commercial high-field / high frequency (HF EPR) and Fourier-Transform EPR spectrometers (FT-EPR) open new horizons in characterization of these materials and their structure elucidation. Some capabilities of the modern EPR spectroscopy to study the HAp based materials are shown in this work. The experimental results are supported by the density functional theory (DFT) calculations.

## 2. Materials and methods

### 2.1. Synthesis of HAp powders

HAp powders with the stoichiometric formula $Ca_{10}(PO_4)_6(OH)_2$ were prepared in the Group of V.I. Putlyaev (Department of the Material Sciances, Moscow State University) by the wet preparation technique according to equation

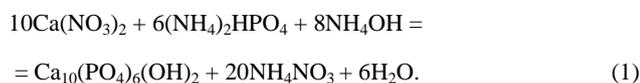

$$= Ca_{10}(PO_4)_6(OH)_2 + 20NH_4NO_3 + 6H_2O. \quad (1)$$

The main advantages of the process are (i) high yield of targeted phase of HAp in one run due to significant solubility of calcium nitrate in water, (ii) fast production rate, (iii) low processing costs, (iv) the only by-product consists of ammonium nitrate irreversibly decomposed at temperature higher than 210°C into two or three gases – water vapor and nitrogen oxide (I), $N_2O$ (below 270°C), or water vapor nitrogen and oxygen (> 270°C), i.e. it is possible, in principle, to obtain HAp of the highest purity after calcination of the precipitate. However, in case of using the HAp precipitate (i) one has to spent time on washing the precipitate to remove nitrate residues and (ii) the resulting product can be greatly affected by even a slight difference in the reaction conditions.

Briefly, 0.3 M stock solution of ammonium phosphate $(NH_4)_2HPO_4$, 3N purity, Labteh, Russia) was added in a dropwise manner under constant stirring in an atmosphere of gaseous $N_2$ (to avoid carbonate contamination) to 0.5 M stock solution containing calcium nitrate $(Ca(NO_3)_2·4H_2O$, 3N purity, Labteh, Russia) at desired ratio. Prior to making the stock solutions all the reagents were checked for the content of main substances by gravimetric or titrimetric chemical analyses. The pH value of the stock solutions was pre-adjusted at $11.0 \pm 0.5$ and then it was maintained manually at the constant value by addition of the concentrated solution of $NH_4OH$. The temperature was controlled and regulated at $80 \pm 1$ °C. After total mixing of the stock solutions, the suspension was ripened and heated by use of a thermostatically controlled hot plate to obtain the micro- and nano-HAp: for 1 hour in one run (the case of 20 nm powder) up to 7 days for 1 μm powder under constant stirring and $N_2$-bubbling. Then, the precipitates were filtered, thoroughly washed with 1 l of distilled water and allowed to dry at room temperature (RT) overnight. Other details of the synthesis and post-synthesis treatments are given in [17, 18].

### 2.2. Analytical methods

The samples were studied by X-ray diffraction in the interval of angles $2\Theta = 10\text{-}110°$ (Cu Kα radiation, Rigaku D/MAX 2500 with rotating anode) and FTIR spectroscopy in 400-4000 cm$^{-1}$ range (Perkin-Elmer 1600). The micromorphology of the powders was examined by scanning and transmission electron microscopy (TEM, JEM-2000FX II, JEOL, operated at 200 kV and FESEM LEO SUPRA 50VP, Carl Zeiss, 5 kV).

The Ca- and P-contents of the synthesized powders were extracted by energy dispersive X-ray fluorescence analysis (EDX, INCA Energy+, Oxford Instruments attached to LEO SUPRA 50VP).

X-ray irradiation of the synthesized nanopowders was performed using URS-55 source (U = 55kV, I = 16 mA, W - anticathode) at RT to create stable paramagnetic centres. XRD patterns before and after X-ray irradiation were checked with Bruker D2 Phaser diffractometer.

FT (pulsed) and continuous wave EPR and ENDOR measurements were done by using X-band (9 GHz) and high-frequency (94 GHz, W-band) Bruker Elexsys 580/680 combined spectrometer equipped with the liquid helium temperature controller and different types of resonators.

EPR spectra were recorded by means of standard cw-technique [15] as well as by means of field-swept two-pulse echo sequence (FS-ESE) $\pi/2 - \tau - \pi$ with the pulse length of $\pi$ pulse of 16 (X-band) or 36 ns (W-band), correspondingly, and time delay $\tau = 240$ ns. For the phase-memory time measurements $\tau$ was varied from 200 ns up to the desired value with the minimal possible step of 4 ns. The Inversion-Recovery pulse sequence $\pi-\tau-\pi/2-\tau-\pi$ for the spin-lattice relaxation times measurements was used. ENDOR spectra were detected by Mims sequence [15, 19] with the length of the radiofrequency (RF) pulse of 18 μs. RF frequency could be swept in the range (1-200) MHz. Simulations of the obtained EPR and ENDOR spectra are done using EasySpin toolbox for MatLab [20].

### 2.3. Density functional theory based calculations

Ab-initio calculations have been carried out within the framework of the plane-wave pseudopotential DFT implemented into the open-source Quantum ESPRESSO package [21]. Modified Perdew-Burke-Ernzerhof version of the generalized gradient approximation of the exchange and correlation functionals (GGA-PBE) [22] together with the Vanderbilt ultrasoft pseudopotentials (USPP) [23] were employed. We have explored the influence of the supercell size on the formation energies of defects and HAp structural parameters. Unit-cell and initial positional parameters for HAp were taken from [24]. Extensive convergence test defined the energy cutoffs of 40 Ry for the smooth part of the electron wave functions and 320 Ry for the augmented electron density. The Brillouin zone was sampled only at the Γ-point whereas the total energies are calculated within an additional self-consistent field (SCF) cycle using 2x2x1 Monkhorst-Pack k-point set [25].

Constants of the hyperfine interaction of the unpaired electron with $^{14}N$ nucleus were extracted by using the gauge-including projector augmented plane wave (GIPAW) approach [26] and Troullier-Martins pseudopotentials [27] with the plane-wave cutoff energy of 70 Ry.

# 3. Results

As X-band EPR measurements show, the synthesized according to Eq. 1 nanopowders of $Ca_{10}(PO_4)_6(OH)_2$ after washing, drying, centrifugation and thermal treatment are EPR silent. No traces of the nitrogen containing impurities are determined neither by EDX (with the beam energy below 10 keV) nor by FTIR spectroscopy applied.

XRD analysis show that all the samples contain only one phase corresponding to the space group $P6_3/m$ with the parameters of the unit cell close to $a \approx b \approx 0.942$ nm and $c = 0.688$ nm typical for the bulk crystals of the hydroxyapatite [6]. No great differences in the positions and integral intensities of XRD peaks were observed with the crystal sizes in the range of (20 nm - 1 μm). Sizes of crystallites were estimated from the diffraction line profile (002) (i.e. from the corresponding line broadening with the size decrease) using Scherrer- and Wilson-formulas from Williamson-Hall-plots according to the procedure described in [28]. Corresponding EDX investigations ensure the ratio Ca/P = 1.66(2) within the accuracy of our measurements in the powder grains serving as an evidence of the stoichiometry of the investigated HAp species.

The samples were X-ray irradiated at RT. XRD patterns before and after X-ray irradiation with the doses of 5 and 10 kGy are measured to be the same. It shows that the crystal structure of the nanopowders does not change under X-ray irradiation with the indicated doses.

The EPR spectrum of X-ray irradiated nano-HAp samples with an average size of 30 nm recorded in cw-mode at RT is presented in Fig. 1A. The cw EPR pattern for all the investigated samples is the same. After X-ray irradiation with the estimated doses of 5 kGy (during 30 minutes), 10 kGy (during 1 hour) and about 100 kGy (overnight ionization) the amplitude of the EPR signal remains the same. Induced radicals' concentration of $5(1)·10^{18}$ spins per gram (about 0.4 mole %) was estimated from the comparison of the integrated intensity of the spectra with that from the reference ($Mn^{2+}$ in MgO). It means that all the species responsible for the EPR spectrum are ionized at 5 kGy and no additional $NO_3^{2-}$ radicals could be created even at higher X-ray doses at RT. We did not investigate the concentration dependence at lower irradiation doses and at lower temperatures.

Under the storage at RT in the laboratory cupboard $NO_3^{2-}$ concentration is dropping down slowly. We have estimated the half lifetime of the radical in our samples as $t_{1/2} \approx 2$ years during the 3 years of measurements.

Usually, the cw EPR spectrum of irradiated bone, tooth enamel or modified HAp contains a multitude of radiation-induced and radiation insensitive signals that can overlap each other complicating, therefore, their interpretation [16]. In this sense the application of the selective pulse methods can simplify the spectra as well as can provide some additional information. A prerequisite to exploiting the variety of the pulsed techniques is an observation of electron spin echo (ESE).

Fig. 1B presents the EPR spectrum obtained by FS-ESE in X-band (ν = 9.6 GHz) at RT along with the simulation for $NO_3^{2-}$ type radical (see section 4 for details). It gives an opportunity to measure characteristics relaxation times of the centers observed. For X-band at T = 298 K spin-lattice (longitudinal) relaxation

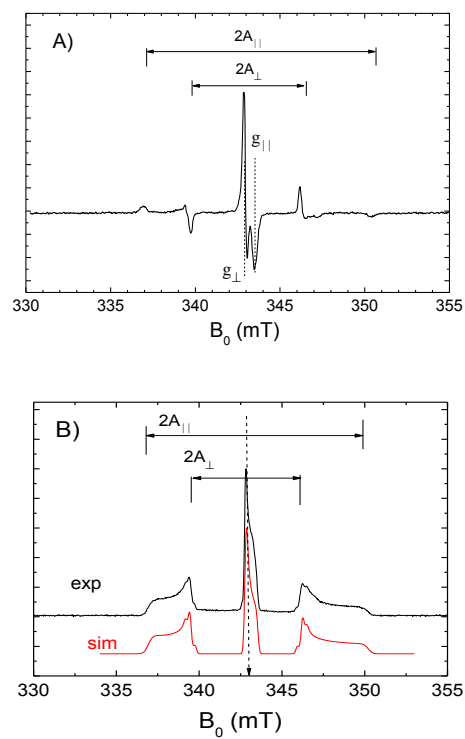

**Fig. 1** EPR spectra in a powder of $Ca_{10}(PO_4)_6(OH)_2$ nanoparticles (30 nm) after X-ray irradiation in X-band at T = 298 K measured in (A) cw-mode and (B) in FS-ESE mode (exp) along with the simulation for $NO_3^{2-}$ radical curve (sim) with the parameters given in Table 1. Simulation curve is vertically shifted for convenience. The vertical dashed arrow points the value of $B_0$ in which ENDOR spectra are acquired and relaxation times are measured

time $T_{1e} = 22(2)$ μs is estimated from the Inversion-Recovery experiments while the phase-memory (transversal) time $T_2^* = 5.3(2)$ μs from the spin-echo decay in the magnetic field $B_0$ corresponding to the perpendicular orientation of the nanocrystals investigated (maximal ESE, vertical dashed arrow in Fig. 1B).

EPR spectra show that at least three modifications of the $NO_3^{2-}$ type radical could be distinguished in the investigated samples. This fact is discussed in section 4. No detectable changes of the X-band EPR spectra are observed with temperature in the temperature range (16-300) K.

In [29] it was proposed that the HAp components of the aorta organomineralic matrix could serve at least as markers for the diagnosis of the atherosclerotic plaque formation/rupture risks. Advantages of the increased resolution and sensitivity of the HF EPR methods in comparison to the conventional X-band were applied to detect and quantify the tissues containing the carbonate radicals in nano-HAp inclusions for the early diagnosis, to investigate spectral and relaxation characteristics of the naturally existing manganese or intentionally incorporated lead ions and to establish their locations in atherosclerotic plaque and nano-HAp structures [18, 30-34]. W-band FS-ESE detected EPR spectrum of the 20 nm sample at T = 50 K is presented in Fig. 2. (The FS-ESE spectrum taken at RT has poor signal-to-noise ratio that significantly increases the measurement time and does not allow detecting all the features obtained). Relaxation times $T_{1e}$ = 30(1) μs and $T_2^*$ = 8.3(3) μs are estimated at T = 300 K in the magnetic field $B_0$ corresponding to the maximal obtained ESE (dashed arrow on Fig. 2). This value of $B_0$ was chosen (i) to reduce the measurement (accumulation) time due to the highest amplitude along the spectrum, (ii) it corresponds to the spectral position which is thought to be extremely separated from other presumably existing paramagnetic centers, and (iii) it corresponds to the spectral position in which the $NO_3^{2-}$ radicals are oriented perpendicular to the $B_0$. Nature of the additional splittings detected in the W-band experiments is discussed in Section 4.

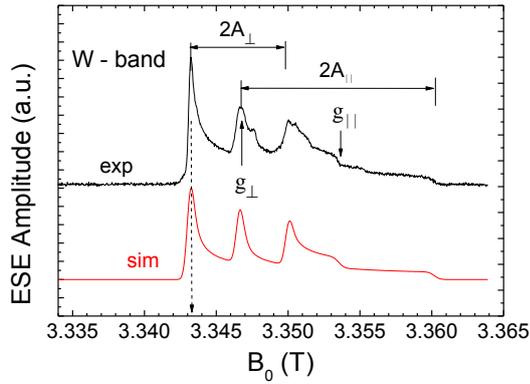

**Fig. 2** FS-ESE EPR spectrum obtained in a powder of nano-HAp (20 nm) after X-ray irradiation in the W-band at T = 50 K (exp). Simulation curve (sim) for the NO32- radical with the same parameters as for the X-band measurements (Table 1) is vertically shifted for convenience. The vertical dashed arrow marks the value of B0 in which the relaxation times and ENDOR spectra are measured. The position corresponding to the g-factor of the free electron is also marked

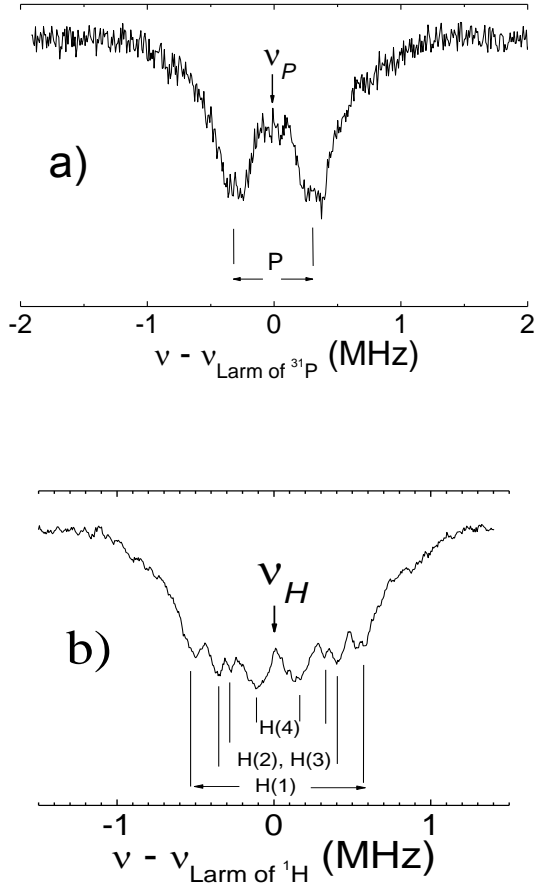

**Fig. 3** ENDOR spectra of the X-ray irradiated nano-HAp in the vicinity of (a) phosphorous and (b) hydrogen Larmor frequencies; W-band, T = 50 K

To specify the coordination of the obtained paramagnetic centers ENDOR experiments were carried out. Fig. 3 presents the Mims-ENDOR spectra in the vicinity of phosphorous ($^{31}P$) and hydrogen ($^1H$) Larmor frequencies obtained in W-band at T = 50 K. The ENDOR spectra at T = 300 K in W-band and at T = 50 K and 300 K in X-band are broader, have low signal-to-noise ratio and are not shown. The interpretation of the observed ENDOR splittings is given in Section 4.

No changes of the observed EPR, ENDOR signals or relaxation times were monitored even after samples annealing up to 400 °C for 4 hours under vacuum (< $10^{-4}$ mbar). Our attempts to replace the observed paramagnetic centers by washing the X-irradiated HAp powders in a NaOH or HCl solutions in a wide pH range (3-12) also do not reveal a significant change in EPR or ENDOR spectra or relaxation characteristics. These results demonstrate that the observed radicals are not located on the powder grain surface but most probably are incorporated into the HAp structure of the investigated HAp nanopowders. It also could be concluded that the proton ENDOR spectrum is mainly due to the superhyperfine interaction with $^1H$ nuclei of HAp rather than with those of the embedded into the HAp structure lattice water molecules.

No significant difference between the EPR and ENDOR signals or relaxation times were observed with the size of the powder grains.

## 4. Discussion

### 4.1. EPR spectra

The dominant feature of the EPR spectra is an intensive three-line signal. The position of the EPR spectra and observation of the three-line pattern with the characteristic splitting (due to the hyperfine interaction between the electronic spin with S = ½ and one nuclear spin with I = 1) indicate the presence of stable nitrogen-containing paramagnetic center (for $^{14}N$ isotope with the natural abundance of 99.63% I = 1). The observed X-band spectra (Fig. 1) present a typical powder spectra that can described by the spin-Hamiltonian of the axial symmetry

$$H = g_\parallel \beta B_z S_z + g_\perp \beta \left(B_x S_x + B_y S_y\right) + A_\parallel S_z I_z + A_\perp \left(S_x I_x + S_y I_y\right), \quad (2)$$

where $g_\parallel$ and $g_\perp$ are the main components of the g-tensor, $A_\parallel$ and $A_\perp$ are the main components of the hyperfine tensor, $B_i$, $S_i$ and $I_i$ are the projections of the external magnetic field strength, electronic spin S = ½ and nuclear spin I = 1, correspondingly, onto the i = {x, y, z} coordinate axis, β is a Bohr magneton.

The values listed in Table 1 are derived from the simulations of the ESE spectrum (Fig. 1B). They show the presence of at least three paramagnetic centers of the same origin (denoted as M for "modification") with slightly different parameters. In the simulations it is assumed that A- and g- tensors are coliniear.

The X-band spectra could be simulated nicely if we suppose a Gaussian (continuous) distribution of $A_\parallel$ with a deviation of 0.40 mT around the mean value of 6.65 mT and discrete distribution of $A_\perp$ around 3.37 mT while the components of $g$-factors for all of the obtained radical's modifications are the same. The most intensive signal from the modification denoted as MI is overlapped by two others in less concentration (Table 1). EPR spectra undoubtedly indicate the presence of the nitrogen containing species that can be used for the purity and affinity check of the hydroxyapatites synthesized for the hard tissues implantation.

**Table 1.** EPR parameters extracted from the simulations of the ESE RT X-band spectra (Fig. 1)

| Modification | $g_\parallel$ | $g_\perp$ | $A_\parallel$, mT | $A_\perp$, mT | Statistical weight |
|---|---|---|---|---|---|
| MI | 2.0011(1) | 2.0052(1) | 6.65(40) | 3.37(5) | > 66 % |
| MII | 2.0011(1) | 2.0052(1) | 6.65(40) | 3.64(5) | < 17 % |
| MIII | 2.0011(1) | 2.0052(1) | 6.65(40) | 3.05(10) | < 17 % |

As it was mentioned, an interest to the (in)organic nitrogen containing radicals is caused by their active and signaling roles in many biological processes [11]. A stable inorganic nitroxyl (NO·) radical Fremy's Salt firstly synthesized more than 160 years ago [34] attracts a great attention nowadays concerning its modern applications as an antioxidant, polarizing agent for dynamic nuclear polarization (DNP), contrast agent in magnetic resonance imaging, as HF EPR probe, etc [37, 38]. New results obtained on this nitroxyl (mainly due to the advanced HF EPR and HF DNP techniques) force to re-examine some of the well-established models for the many aspects of the nitroxyl radicals' dynamics and interactions with the surrounding. Peroxynitrite anion ONOO- is considered as one of the most reactive nitrogen species that can damage a wide range of molecules in cells including DNA [39]. Another point of interest is connected with $NO_3$ radical - an important nighttime oxidant in the Earth's atmosphere. Firstly observed over 130 years ago, this radical is an example of the breakdown of the Born-Oppenheimer approximation which absorption spectrum, electronic structure, intermolecular dynamics and photochemistry are still not fully understood and new theoretical and experimental approaches are introduced for their study [40, 41].

Different inorganic radicals trapped in natural and synthesized crystals could be detected and separated by EPR. The most stable among them are of the general type of $RO_2$ and $RO_3$ (R is the central atom, O is an oxygen) [42]. After hot discussions about the assignments of the observed by different authors EPR spectra to various radicals it seems to be that since end of 1960s a consensus is achieved. A large number of the EPR experiments supported by theoretical calculations on oriented inorganic free radicals carried out since 1950s show that generally the most $RO_2$ type radicals, (like $NO_2$, $NO_2^{2-}$) have anisotropic hyperfine interaction and $g$-tensors with all three principal values unequal. The $RO_3$ type radicals ($NO_3$, $NO_3^{2-}$), on the other hand, have an axially symmetric hyperfine interaction tensor and (in the most cases) $g$-factor [41-52]. It could be expected also that due to the additional rotational mobility of $RO_2$ and OROO type radicals along O-R-O axis [50, 50] their electronic spin-lattice relaxation times should be shorter than those for $RO_3$ and it might be problematic to obtain ESE detected EPR of those in HAp at RT. Lists of $g$- and $A$- values for some of the stable nitrogen-centered radicals are gathered in the Landolt-Börnstein database [53, 54]. A part of them of interest are presented in Table 2 to show their typical values.

**Table 2.** Some literature data of the EPR parameters (main components of $g$- and $A$-tensors) for some of the stable inorganic nitrogen-centred radicals

| Radical | Matrix/Conditions | $g$-factor | $A$-values ($^{14}$N), mT | Reference |
|---|---|---|---|---|
| $NO_2$ | CaX zeolite, 114 K | 2.0051, 1.9921, 2.0017 | 5.19; 4.78; 5.58 | [55] |
| $NO_2^{2-}$ | Apatite, 77K | 2.0014; 2.0067; 2.0051 | 0.4; 0.5; 3.88 | [50] |
| $NO_3$ | KNO3 single crystal γ-irr EPR / 77 | 2.0075; 2.0075; 2.0018 | 0.06; 0.12; 0.22 | [53, 54] and references therein |
| $NO_3^{2-}$ | HAp, X-irr, RT | 2.006; 2.006; 2.0019 | 3.4; 3.4; 6.8 | [56] |
|  | HAp, X-irr, RT | 2.0061, 2.0061, 2.0028, | 3.18; 3.18; 6.83 | [57] |
|  | Nano-HAp, X-irr., 50-300 K. | 2.0052; 2.0052; 2.0011 | 3.35(35); 3.35(35); 6.65(40) | This work |
| ONOO | KNO3 single crystal γ-irr EPR / 77 | 2.0029; 2.0203; 2.0203 | 0.36; 0.38; 0.06 | [45] |
| NO (Fremy Salt) | Crystals, frozen solutions | 2.0081; 2.0057; 2.0025 | 0.55; 0.4; 2.9 | [26] and references therein |

Some researchers still incorrectly interpret the similar to Fig. 1A EPR spectra ascribing them to other types of (carbonated) radicals due to the hydrogen hyperfine splitting, for example [58, 59]. This fallacy is caused among other things by the absence of the nitrate bands in the vicinity of 825 and 1385 cm$^{-1}$ in IR patterns [60]. The last sometimes (also in our measurements) are observed in the fresh prepared species before the washing and centrifugation [60]. They could be caused by the surface adsorbed species or by the nitrates in mother solution. Their origins are usually assigned to the $v_2$ (out-if-plane bending) and $v_3$ (degenerate N-O stretch) modes of $NO_3^-$ in $D_{3h}$ symmetry, correspondingly [41], though their positions and nature are still under discussion [40]. The lack of the nitrate infrared response could be ascribed to the low nitrate concentration, IR lines broadening with the particle size decrease or to their overlap with the intensive, broad $PO_4^{3-}$ bands. It seems to be that some special measurement protocols should be applied to detect the nitrate residues in nano-HAp by IR spectroscopy [40]. In this sense the exploited EPR routines could serve as a sensitive supplement to the recognized methods of the nitrate detection [61].

Based on the values of *g*- and *A*- tensors obtained in our experiments we can conclude that the X-band and W-band EPR spectra are undoubtedly due to the presence of $NO_3^{2-}$ type radical.

Two modifications of $NO_3^{2-}$ radicals having slightly different $A_\perp$ with the parameters close to MI and MII obtained in our experiments were registered by I.P. Vorona et al. in bulk HAp only under UV-irradiation by means of cw EPR [46, 47]. These authors also obtained another $NO_3^{2-}$ modification with $A_\perp$ up to 4.41 mT which appear under different pre-annealing treatment of the materials investigated [51]. They observed that though the modification MI (in our notations) is a dominant one, after annealing even at 200 °C the intensities of EPR signals of MI and MII drop down while the observed value of $A_\perp$ of MI grows up. It was concluded that the modifications are caused either by the presence of the paramagnetic crystal defects (impurities) that reduce the electronic spin density on the nitrogen nuclei or by different radical location in the matrix. Following the cited above papers, we can further conclude that the distribution ratio between the different modifications could be in principal used to specify the quality of the synthesized HAp materials or to estimate the radical distribution between the location sites.

In contrast to the cited works [46, 51] we have managed to detect one more $NO_3^{2-}$ (MIII) modification with even smaller value of $A_\perp$ in X-ray irradiated material and did not observe great changes in the EPR spectra after annealing up to 400 °C in our nano-HAp samples. It seems to be that the variety of the modifications obtained in HAp is caused by the quality of the samples rather than by the type of the radiation source and its dose. It also be noted that it is quite unusual situation that paramagnetic defects could increase the values of $A_\perp$ or $A_\parallel$ as we

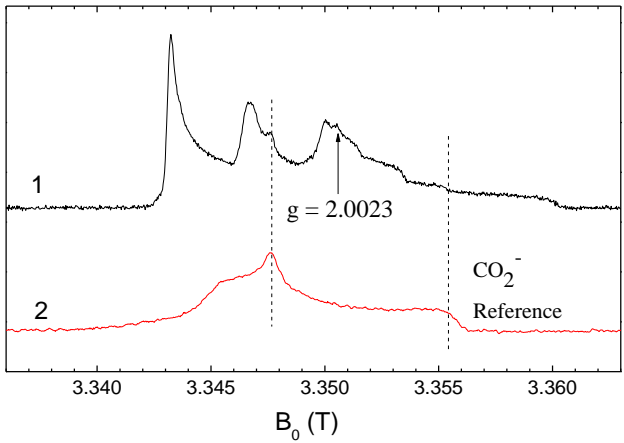

**Fig. 4** W-band FS-ESE EPR spectra of (1) X-ray irradiated nano-HAp and (2) $CO_2^-$ radical from the reference material (the mammoth tusk). Arrow shows the position in the spectrum corresponding to the *g*-factor of the free electron

observe in our experiments (modification MII).

The amazing stability of the radical and its sufficiently long relaxation times force us at least to pay more attention to their (electronic) structure. Following Walsh [48], Atkins and Symons [49] one expects the resulting $NO_3^{2-}$ to be pyramidal with the $C_{3v}$ symmetry since in this case the extra electron is thereby stabilized: the 25$^{th}$ unpaired electron upon entering the antibonding $2a_2$ orbital causes bending of the 24-electron planar structure of the nitrate anion $NO_3^-$. The distortion introduces some *s* character to this orbital reducing its antibonding and both isotropic and anisotropic hyperfine contributions from the central N atom are expected leading to the large values of the *A*-components and their anisotropy in comparison to other radicals. The out of plane deviation is usually unknown and estimated to be in the range (10-16)°. The NO bond lengths of the Y-shaped nitrate radicals are estimated to be of 0.124 nm for $NO_3$ up to 0.127 nm for $NO_3^{2-}$. No deviations from the axial symmetry were observed even in single crystals by X-band EPR [52].

Therefore, we are not allowed to ascribe the additional splittings observed in the W-band experiments (Fig. 2) to the rhombicity of the *g*-factor though formally it would be possible to simulate the whole W-band spectrum in this manner. The possibility to detect additional radiation induced EPR signal(s) at high EPR frequency could be ascribed to the increased spectral resolution and sensitivity of the HF EPR as well as to the changes of the relaxation times of the hypothesized paramagnetic centers with $B_0$. We ascribe the obtained additional peaks to the superposition of the EPR spectrum of a carbonate radical (presumably, air absorbed) and spectrum of some paramagnetic centre(s) with the spectroscopic *g*-factor close to that of the free electron. EPR of an X-ray irradiated powder of mammoth tusk served us as a reference for the spectroscopic parameters of the carbonate radical (Fig. 4). The detailed EPR and ENDOR investigation of carbonation of the synthesized HAp samples is a matter of the ongoing research.

### 4.2. ENDOR spectra

Superhyperfine interaction, characterized by the superhyperfine constant *a*, between the electron cloud of $NO_3^{2-}$ radical and neighbouring nuclei with $I = 1/2$ like $^{31}P$ and $^1H$ leads to the splitting of the ENDOR spectra according to

$$\nu_{ENDOR} = h^{-1}|g^{(I)}\beta^{(I)}B_0 \pm a/2|, \qquad (3)$$

where *h* is a Planck constant, $g^{(I)}$ is a nuclear *g*-factor and $\beta^{(I)}$ is a nuclear Bohr magneton [19].

For the first approximation, assuming the pure electron-nuclei dipole-dipole interaction in the point model, the electron-nuclear distances, *r*, from the ENDOR splitting can be estimated:

$$a \propto g \cdot g^{(I)} (1-3\cos^2\Theta)/r^3, \qquad (4)$$

where $\Theta$ is an angle between and directions of $g_\parallel$ and $B_0$. Spectral resolution of the W-band experiments allows to choose the value of $B_0$ for ENDOR corresponding to the "pure" perpendicular orientation $g_\perp$. In this case the factor $(1-3\cos^2\Theta) = 1$. The calculations are registered as $r_{exp}$ in Table 3. One phosphorous and different hydrogen nuclei neighbouring the radical could be distinguished in our experiments (see Fig. 3). Due to the closeness of the spectroscopic parameters of all three modifications MI-MIII to each other (cf. Table 1), it is expected that all of them give an equal contribution to the ENDOR spectra.

Table 3. ENDOR splittings, electron-nuclear distances extracted from the ENDOR experiments using Eq. 4 in the point-point dipole-dipole interaction approach ($r_{exp}$), distances extracted from the crystal structure of HAp [6] between the atom in one substituted position in OH channel and two substituted $PO_4^{3-}$ positions and the nearest $^1$H and $^{31}$P nuclei ($r_{calc}$) and total deviation between $r_{exp}$ and $r_{calc}$.

|  | ENDOR splitting (MHz) | | | $r_{exp}$ from W-band 50 K (nm) | $r_{calc}$ from the indicated positions to the nearest nuclei in the HAp structure (nm) | | |
| --- | --- | --- | --- | --- | --- | --- | --- |
|  | X-band, RT | W-band, RT | W-band, 50 K |  | OH channel, position N(1) | $PO_4^{3-}$ position, N(2) | $PO_4^{3-}$ position, N(3) |
| H(1) | 0.66(20) | 0.78(12) | 1.06(1) | 0.42(1) | 0.32 | 0.48 | 0.6 |
| H(2) |  |  | 0.75(1) | 0.47(1) | 0.37 | 0.5 | 0.61 |
| H(3) |  |  | 0.60(1) | 0.51(1) | 0.66 | 0.67 | 0.76 |
| H(4) |  |  | 0.26(1) | 0.67(1) | 0.71 | 0.71 | 0.8 |
| P | 0.72(25) | 0.70(20) | 0.64 (6) | 0.37(1) | 0.40 | 0.32 | 0.4 |
|  |  |  |  | $\Sigma(r_{exp}-r_{calc})^2$ | 0.045 nm$^2$ | 0.034 nm$^2$ | 0.132 nm$^2$ |

Hydroxyapatite structure offers a variety of opportunities for incorporation and substitution of different ions, atoms and structures. HAp structure could be considered as formed by a tetrahedral arrangement of phosphate ($PO_4^{3-}$) which constitute the "skeleton" of the unit cell. Two of the oxygen atoms are aligned with the *c*-axis and the other two are in a horizontal plane [6]. Within the unit cell, phosphates are divided into two layers, with heights of 1/4 and 3/4, respectively, resulting in the formation of two types of channels along the *c*-axis, denoted by A (or OH-channel) and B (or Ca(1) channel). The walls of channels of the A-type are occupied by oxygen atoms of phosphate group and calcium ions, called calcium ions type II [Ca(2)]. Type B channels are occupied by other ions of calcium, called calcium ions type I [Ca(1)]. In the stoichiometric HAp the centers of the channels type A are occupied by OH radicals, with alternating orientations.

The Ca sites are both available for cationic substitution while carbonates, for example (which content varies in the range (3-8) wt % in the human calcified tissues), can substitute both the hydroxyl and the phosphate ions, giving rise to the A-type and B-type carbonation, respectively [16]. The knowledge about the location of the substituent could help to understand the mechanisms of their accumulation in tissues of the body, to differentiate the functions of the channels in biological processes, to check the affinity of the synthesized HAp to bone tissue, to follow the metabolic processes within the host organisms, *etc*. For example, ENDOR investigations of Pb-containing nano-HAp synthesized by the same wet procedure as in the present paper demonstrate straightforwardly that the divalent lead ions are at least distributed between Ca(1) and Ca(2) and $Pb^{3+}$ ions replaces Ca(1) while the majority of the experimental data supported by semi-empirical and first-principle calculations show an energy preference for Ca(2) site in micro-HAp [18]. Unfortunately, the ENDOR spectra of $NO_3^{2-}$ do not give such categorical answer to a radical location question.

Since $NO_3^-$ and $CO_3^{2-}$ are isoelectronic ions with the same symmetry of ground sate ($D_{3h}$) and close ionic radii, it is logically to assume that $NO_3^-$ occupy the same sites that carbonate does [41]. But, as it was mentioned above, depending on the preparation procedure or type of the natural HAp containing material carbonates can be found in different locations [62]. No traces of the carbonate containing species were revealed by IR and X-band EPR in our samples. Low concentration of the presumable $CO_2^-$ centre and its spectral overlapping with other signals in W-band does not allow defining its position by ENDOR.

We have considered three positions for the nitrogen atom of the nitrate ion: N(1) that substitutes two close hydrogen atoms in OH-channel; N(2) and N(3) that substitute O and P atoms and phosphate tetrahedron, correspondingly (Fig. 5). The distances between these positions and nearest hydrogen and phosphorous atoms ($r_{calc}$), derived from the undistorted perfect HAp crystal structure are listed in Table 3.

In this simple primitive model two positions N(2) and N(1) appear to be the most probable for the nitrate incorporation (as in the case of carbonate substitution they can be designated as B- and A-sites, respectively). From this analysis it follows that in the investigated samples the nitrate ions could be distributed between those two. However, bearing in mind excellent stability of $NO_3^{2-}$ we can speculate that pyramidal symmetry of the radical fits rather tetrahedral B-sites than A-sites with axial symmetry.

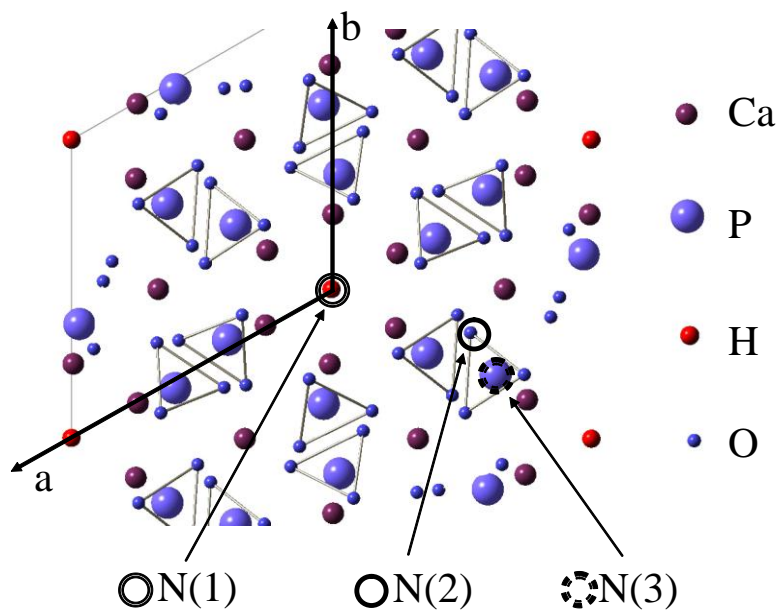

**Fig. 5** Proposed positions of the nitrogen atoms N(1), N(2) and N(3) of $NO_3^{2-}$ radicals in the HAp structure in simplified form (cross section is in the *ab* plane). The sticks designate the edges of $PO_4^{3-}$ tetrahedral. The size of the hydrogen atoms (nuclei) are intentionally presented larger than that of oxygen as a guide for eyes because the hydrogen splitting are to observe in ENDOR experiments.

### 4.3. DFT calculations

We have considered two possible locations of $NO_3^-$ ion in the HAp lattice: N(1) and N(2) as followed from the ENDOR data (see Fig. 5). For the last one a charge compensation is required. As it is usually assumed for the carbonate-doped HAp, we suggest that the charge compensation scheme is manifested in the removal of one of the nearest Ca [63]. We have performed full geometry optimization runs with different initial configurations. Namely, for the B-site location we set the impurity ion so that the plane of $NO_3^-$ was either parallel or perpendicular to the crystallographic *b*-axis at different heights with respect to Ca-plane; for $NO_3^-$ in A-site both parallel and perpendicular initial orientations with respect to *b*-axis were regarded.

As it follows from the computations, for both of the defect models the substitutional nitrate ion has almost identical planar configuration with the average N-O distance of about 0.126 nm. The lattice parameters for both of the simulated models are appearing to be very close to those for pure HAp (the deviations do not exceed 1 %). No A- or B-site location preference for $NO_3^-$ are predetermined.

In order to study the structure and magnetic properties of the paramagnetic $NO_3^{2-}$ centre we have performed spin-polarized GGA-PBE calculations by adding an extra electron. The structural relaxation is carried out with respect to atomic positions with the lattice parameters fixed. The optimized supercells containing $NO_3^-$ impurity are used as initial data. The structure of the resulting $NO_3^{2-}$ radical show significant deviation from the planarity for both B- and A-sites towards to the pyramidal shape. The out of plane deviation is estimated to be of about 13°. The values of the spin density localization on $^{14}N$ are computed to be of 0.29 for the B-site and 0.37 for the A-site, correspondingly. The hyperfine tensor components are presented in Table 4. From the excellent correspondence with the experimental results, it can be concluded that the obtained EPR and ENDOR spectra originate from the B-site located radicals only.

**Table 4.** DFT calculated $^{14}N$ hyperfine components for the A- (OH channel) and B-($PO_4^{3-}$) types of substitutions (in mT)

|  | $A_x$ | $A_y$ | $A_z$ |
|---|---|---|---|
| A-site, N(1) | 2.485 | 2.478 | 5.183 |
| B-site, N(2) | 3.277 | 3.273 | 6.413 |

Based on that conclusion, we can further assume that the exsistence of different $NO_3^{2-}$ radical modifications obtained in our experiments (cf. Table 1) is probably due to the small deviations in their arrngements in the B-site rather than due to the radical space distribution along the powder grain (we see no great difference in the experiment with the crystalites size). From the independence of our EPR experimental results on the radiation dose, on the attempt to replace the nitrates by OH-ions, etc. we can speculate that $NO_3^-$ as precursors for the paramagnetic radicals are also probably occupy B-site.

### 4.4. Relaxation Times

The similarity of the relaxation times of the observed centers in (20-50) nm and 1 μm samples can serve as an additional proof that the obtained EPR signals are mainly originated from the incorporated rather than from the surface radicals. The closeness of the relaxation times to each other in different magnetic fields at RT (≈ 0.35 T for X-band and ≈ 3.4 T for W-band) is typical for the Raman type relaxation mechanisms in crystals [64, 65]. Sufficiently long $T_2^*$ is a sign of the homogeneous space distribution of the radicals in the sample volume and indicates the absence of other near-located paramagnetic centres. It is known that biological HAp differs from the synthetic HAp in several respects, including non-stoichometry and poor crystallinity [2, 14]. In this sense, the relaxation times of $NO_3^{2-}$ could be considered as the additional parameters qualifying the micro- and nano-HAp crystallinity and quality.

## 5. Conclusion and perspectives

Modern EPR approaches can complement the standard analytical tools for the comprehensive characterization of a HAp based materials [60] and could be used for the quality check, estimations of the crystallinity and purity of the synthesized species and for the further structural investigations of the HAp based materials.

$NO_3^{2-}$ radicals with slightly different spectroscopic parameters in the wet-synthesized HAp samples are detected by FT-EPR. An increased spectral resolution of HF EPR allows detecting different paramagnetic centers. Sufficiently long relaxation times and observation of ESE and ENDOR at room temperatures give an opportunity to investigate the obtained nitrogen-containing centers by other advanced magnetic resonance methods such as electron-electron double resonance (ELDOR or DEER) to study the radical distribution, its interaction with other paramagnetic impurities and dopants and follow the dynamics in the temperature measurements.

We hope that the presented work could serve as an another node for the detailed investigation of the nitrogen-containing complexes and their dynamics. All these, finally, could help to synthesize the materials with the increased biocompatibility and higher efficiency for the removal of different toxic elements.

## References

1. LeGeros RZ. Properties of osteoconductive biomaterials: calcium phosphates. Clin Orthop Relat R 2002; 395 : 81-98.
2. Zhou H, Lee J. Nanoscale hydroxyapatite particles for bone tissue engineering. Acta Biomater 2011; 7: 2769-2781.
3. Loo SCJ, Moore T, Banik B, Alexis F. Biomedical applications of hydroxyapatite nanoparticles Curr Pharm Biotechnol 2010 ; 11 : 333-342.
4. Uskoković V, Uskoković DP. Nanosized hydroxyapatite and other calcium phosphates: chemistry of formation and application as drug and gene delivery agents. J Biomed Mater Res A 2011; 96B: 152-91.
5. Chu SH, Feng DF, Ma YB Li ZQ. Hydroxyapatite nanoparticles inhibit the growth of human glioma cells in vitro and in vivo. Int J Nanomedicine 2012; 7: 3659-3666.
6. Elliott JC. Structure and chemistry of the apatites and other calcium orthophosphates.Stud Inorg Chem 1994; 18: 1-389.
7. Ma QY, Traina SJ, Logan TJ Ryan JA. In situ lead imobilization by apatite. Environ Sci Technol 1993; 27: 1803-1810.
8. Mizuno N, editor. Modern heterogeneous oxidation catalysis: design, reactions and characterization. Weinheim: Wiley-VCH; 2009. 356 p.
9. Jarcho M, Bolen CH, Thomas MB, Bobick J, Kay JF, Doremus RH. Hydroxyapatite synthesis and characterization in dense polycrystalline form. J Mater Sci 1976; 11: 2027-2035.
10. Kehoe S. Optimisation of hydroxyapatite (HAp) for orthopaedic application via the chemical precipitation technique. Ph.D. Thesis, Dublin: City University; 2008. 393 p.
11. Valko M, Leibfritz D, Moncol J, Cronin MT, Mazur M, Telser J. Free radicals and antioxidants in normal physiological functions and human disease Int J Biochem Cell Biol. 2007; 39: 44-84.
12. Hukkanen MVJ, Polak JM Hughes SPF. Nitric oxide in bone and joint disease. Cambridge: Cambridge University Press; 1998. 191 p.
13. Shi ZL, Huang X, Cai YR, Tang RK, Yang DS. Size effect of hydroxyapatite nanoparticles on proliferation and apoptosis of osteoblast-like cells. Acta Biomaterialia 2009; 5: 338–345.
14. Dorozhkin SV. Nanosized and nanocrystalline calcium orthophosphates. Acta Biomaterialia 2010; 6: 715-734.
15. Weil JA, Bolton JR. Electron paramagnetic resonance: elementary theory and practical applications. 2nd edition. Hoboken, New Jersey: John Wiley & Sons; 2004. 664 p.
16. Fattibene P, Callens F. EPR dosimetry with tooth enamel: a review. Appl Radiat Isot 2010; 68: 2033–2116.
17. Kovaleva ES, Shabanov MP, Putlyaev VI, Tretyakov VD, Ivanov VK,. Silkin NI. Bioresorbable carbonated hydroxyapatite Ca10-xNax(PO4)6-x (CO3)x(OH)2. Cent Eur J Chem 2009; 7: 168-174.
18. Yavkin BV. et al. Pb3+ radiation defects in Ca9Pb(PO4)6(OH)2 hydroxyapatite nanoparticles studied by high-field (W-band) EPR and ENDOR Phys Chem Chem Phys 2012; 14 : 2246-2249.
19. Murphy DM,. Farley RD. Principles and applications of ENDOR spectroscopy for structure determination in solution and disordered matrices. Chem Soc Rev 2006; 35: 249-268.
20. Stoll S, Schweiger A. EasySpin, a comprehensive software package for spectral simulation and analysis in EPR. J Magn Reson 2006; 178: 42-55.
21. Giannozzi P. at al. QUANTUM ESPRESSO: a modular and open-source software project for quantum simulations of materials. J Phys Condens Matter 2009; 21: 395502 (19pp).
22. Perdew JP, Burke K, Ernzerhof M. Generalized gradient approximation made simple. Phys Rev Lett 1996; 77 : 3865-3868.
23. Vanderbilt D. Soft self-consistent pseudopotentials in a generalized eigenvalue formalism. Phys Rev B 1990; 41: 7892-7895.
24. Yashima M, Yonehara Y, Fujimori H. Experimental visualization of chemical bonding and structural disorder in hydroxyapatite through charge and nuclear-density analysis. J Phys Chem C 2011; 115: 25077-25087.
25. Monkhorst HJ, Pack JD. Special points for Brillouin-zone integrations. Phys Rev B 1976; 13: 5188-5192.
26. Pickard C J, Mauri F. All-electron magnetic response with pseudopotentials: NMR chemical shifts. Phys Rev B 2001; 63: 245101 (13pp).
27. Troullier N, Martins JL. Efficient pseudopotentials for plane-wave calculations. Phys Rev B 1991; 43: 1993-2006.
28. De Keijser TH, Langford JI, Mittemeijer EJ, Vogels ABP., Use of the Voigt function in a single-line method for the analysis of X-ray diffraction line broadening. J Appl Cryst 1982; 15: 308-314.
29. Abdul'yanov VA et al. Stationary and high-frequency pulsed electron paramagnetic resonance of a calcified atherosclerotic plaque. JETP Lett 2008; 88: 69-73.
30. Gafurov MR et al. Atherosclerotic plaque and hydroxyapatite nanostructures studied by high-frequency EPR Magn Reson Solids 2013; 15: 13102(7pp).
31. Lozhkin AP, et al. Manganese in atherogenesis: Detection, origin, and a role. Biochemistry (Moscow) Supplement Series B: Biomedical Chemistry 2011; 5: 158-162.
32. Biktagirov TB, Chelyshev YA, Gafurov MR, Mamin GV, Orlinskii SB, Osin YN, Salakhov MK. Investigation of atherosclerotic plaque by high-frequency EPR. J Phys Conf Ser 2013; 478: p.012002.
33. Yavkin BV, Gafurov MR, Kharintsev SS, Mamin GV, Goovaerts E, Salakhov MKh, Osin YuN, Orlinskii SB.

Perspective of zero-field ODMR to study nano-biological systems. J Phys Conf Ser 2013; 478: p.012001.
34. Silkin NI, Chelyshev YuA, Mamin GV, Orlinskii SB, Salakhov MKh. A method for detection of manganese paramagnetic complexes as markers of atherosclerosis. RF Patent №2468368 2012. (in Russian)
35. Silkin NI, Salakhov MKh, Orlinskii SB, Mamin GV, Chelyshev YuA, Galiullina LF, Tokarev GA, Igumnov ES, Gafurov MR. A method for determination of substitutional nitrogen species in hydroxiapatite. RF Patent №2465573 2012. (in Russian)
36. Frémy E. Sur une nouvelle série d'acides formés d'oxygène, de soufre, d'hydrogène et d'azote. Ann Chim Phys 1845; 15: 408-488 (in French).
37. Likhtenshtein G, Yamauchi J, Nakatsuji Sh. Smirnov AI, Tamura R. Nitroxides: applications in chemistry, biomedicine and materials science. Weinheim: Wiley-VCH; 2008. 419 p.
38. Gafurov M, Denysenkov V, Prandolini MJ, Prisner TF. Temperature dependence of the proton Overhauser DNP enhancements on aqueous solutions of Fremy's salt measured in a magnetic field of 9.2 T. Appl Magn Reson 2012; 43: 119-128.
39. Valko M, Leibfritz D, Moncol J, Cronin MT, Mazur M, Telser J. Free radicals and antioxidants in normal physiological functions and human disease. Int J Biochem Cell Biol 2007; 39:44-84.
40. Stanton JF. On the vibronic level structure in the NO3 radical. I. The ground electronic state. Journ Chem Phys 2007; 126: 134309 (20 pp).
41. Wayne RP et al. The nitrate radical: physics, chemistry and the atmosphere. Atmos Environ 1991; 25A: 1-203.
42. Morton JR. Electron spin resonance spectra of oriented radicals. Chem Rev 1964; 64: 453-471.
43. Eachus RS, Symons MCR. Unstable intermediates. Part L. The $NO_3{}^{2-}$ impurity centre in irradiated calcium carbonate. J Chem Soc A 1968; 790-793.
44. Peckauskas RA, Pullman I. Nitrate radicals in apatites. J Dent Res 1975; 54: 1096.
45. Bannov SI, Nevostruev VA. Formation and properties of $NO_3{}^{2-}$, $NO_3$ and $ONOO$ radicals in nitrate-containing matrice. Radiat Phys Chem 2003; 68: 917-924.
46. Vorona IP, Ishchenko SS, Baran NP, Rudko VV, Zatovskiĭ IV, Gorodilova NA, Povarchuk VY. $NO_3{}^{2-}$ centres in synthetic hydroxyapatite. Phys Solid State 2010; 52: 2364-2368.
47. Baran NP, Vorona IP, Ishchenko SS, Nosenko VV, Zatovskiĭ IV, Gorodilova NA, Povarchuk VY. $NO_3{}^{2-}$ and $CO_2{}^-$ centers in synthetic hydroxyapatite: Features of the formation under γ- and UV-irradiations. Phys Solid State 2011; 53: 1891-1894.
48. Walsh D. The electronic orbitals, shapes, and spectra of polyatomic molecules. Part V. Tetratomic non-hydride molecules, $AB_3$. J Chem Soc 1953: 2301-2306.
49. Atkins PW, Symons MCR. The structure of inorganic radicals. Amsterdam-New York: Elsevier; 1967. 280 p.
50. Dugas J, Bejjaji B, Sayah D, Trombe JC. Etude par RPE de l'Ion $NO_2{}^{-2}$ dans une apatite nitrée. Journ Solid State Chem 1978; 24: 143-151 (in French).
51. Nosenko VV, Vorona IP, Ishchenko SS, Baran NP, Zatovsky IV, Gorodilova NA, Povarchuk VY. Effect of pre-annealing on $NO_3{}^{2-}$ centers in synthetic hydroxyapatite. Radiat Meas 2012; 47: 970-973.
52. Canniere PI, Debuyst R, Dejehet F, Apers D. Esr Study of internally α-irradiated (210Po nitrate doped) calcite single crystal. Nucl Tracks Radiat Meas 1998; 14: 267-273.
53. Morton JR, Preston KF. 1.2.6 Nitrogen-centered radicals: in Fischer H, Hellwege KH. (ed.) Landolt-Börnstein – Group II Molecules and Radicals 9A, Berlin-Heidelberg: Springer; 1977. p 50-80.
54. Claridge RFC. 1.7 Nitrogen-centered radicals. in Fischer H, editor. Landolt-Börnstein – Group II Molecules and Radicals Database New series II/26A1. Berlin-Heidelberg: Springer; 2007. p 44-47.
55. Pietrzak TM, Wood DE. EPR study of the hindered motion of $NO_2$ and $ClO_2$ adsorbed in synthetic zeolites. J Chem Phys 1970; 53: 2454-2459.
56. Murata T, Shiraishi K, Ebina Y; Miki T. An ESR study of defects in irradiated hydroxyapatite. Appl Radiat Isot 1996; 47: 1527-1531.
57. Brik AB et al. Nitrogen-containing ion-radicals in biogenic and synthetic calcium phosphates. Miner J (Ukraine) 2006; 28: 20-31 (in Russian).
58. Rokhmistrov DV, Nikolov OT, Gorobchenko OA, Loza KI. Study of structure of calcium phosphate materials by means of electron spin resonance. Appl Radiat Isot 2012; 70: 2621-2626.
59. Schramm D, Rossi A. EPR and ENDOR studies on $CO_2{}^-$ radicals in γ-irradiated B-type carbonated apatites. Phys Chem Chem Phys 1999; 1: 2007-2012.
60. Palard M, Champion E, Foucaud S. Synthesis of silicated hydroxyapatite $Ca_{10}(PO_4)_{6-x}(SiO_4)_x(OH)_{2+x}$ Journ Solid State Chem 2008; 181: 1950–1960.
61. Moorcroft MJ, Davis J, Compton RG. Detection and determination of nitrate and nitrite: a review. Talanta 2001; 54: 785-1020.
62. Gibson R, Bonfield W. Novel synthesis and characterization of an AB-type carbonate-substituted hydroxyapatite J Biomed Mater Res 2002; 59: 697–708.
63. Astala R, Stott MJ. First principles investigation of mineral component of bone: $CO_3$ substitutions in hydroxyapatite. Chem Mater 2005: 17; 4125-4133.
64. Abragam A, Bleaney B. Electron paramagnetic resonance of transition ions. Oxford: Clarendon; 1970. 573 p.
65. Gafurov MR, Aminov LK, Kurkin IN, Izotov VV. Temperature dependence of the EPR linewidth of $Yb^{3+}$ ions in $Y_{0.99}Yb_{0.01}Ba_2Cu_3O_X$ ($6 \leq X \leq 7$) compounds: Evidence for an anomaly near the superconducting transition. Supercond Sci Technol 2005; 18: 352-355.
66. Markovic M, Fowler BO, Tung MS. Preparation and Comprehensive Characterization of a Calcium Hydroxyapatite Reference Material. J Res Natl Inst Stand Technol 2004; 109: 553-568.